# Energy Efficient IoT Virtualization with Passive Optical Access Network

Zaineb T. Al-Azez, Ahmed Q. Lawey, Taisir E.H. El-Gorashi, Jaafar M.H. Elmirghani

*Abstract*— In this paper, the energy efficiency of edge computing platforms for IoT networks connected to a passive optical network (PON) is investigated. We have developed a mixed integer linear programming (MILP) optimization model, which optimizes the placement and number of the cloudlets and VMs and utilizes energy efficient routes with the objective of minimizing the total IoT network and processing power consumption. Our results show that the power consumption can be reduced by consolidating the placement of these VMs at the PON Optical Line Terminal (OLT) in cases where the traffic volume is still high after data processing, ie at low traffic reduction percentages. On the other hand, at high traffic reduction ratios, better power efficiency can be accomplished by placing VMs in lower layer nodes (relays). Our results indicate that utilizing PONs and serving heterogeneous VMs can save up to 19% of the total power. Based on the MILP model insights, a heuristic is developed with very comparable MILP-heuristic power consumption values. We considered three scenarios that represent different levels of homogeneous and heterogeneous VM CPU demands. Good agreement was observed between the heuristic results (17% power saving) and the MILP which results in 19% power saving.

## I. Introduction

As a result of the exponential growth of the Internet traffic, the $CO_2$ emissions and energy consumption of information and communication technology (ICT) networks are undergoing a dramatic increase. This increase is one of the significant challenges that may hinder the expansion of the Internet. Moreover, ICT generates an estimated 2% of the global $CO_2$ emissions [1]. Consequently, more attention must be given to improving energy efficiency and sustainability of the Internet and the ICT industries.

IoT represents a major evolution in legacy data communication. It is predicted that there will be 75 billion IoT interconnected devices by 2025 [2]. This growing level of connected devices has paved the way for futuristic smart applications in healthcare, agriculture, transportation, manufacturing, smart homes and machine-to-machine (M2M) communications [3], [4]. There are, however, many key challenges such as reliability, security, interoperability and scalability [5]. In addition, one of the main challenges that must be confronted by IoT architects is the energy efficiency and greening the networks [6], which is currently garnering attention in both the academic and the industrial arenas. IoT is also expected to benefit from the wide spectrum of proposed energy efficient network solutions. Cloud computing was investigated as one of the solutions that can improve the utility of IoT by storing and processing the IoT generated data. The energy efficiency of cloud data centres was investigated in [7]-[11]. Virtualization can help improve resource sharing in IoT networks and in the supporting data centres and networks, and this was evaluated in [12]-[14]. IoT nodes are typically connected to the access layer of the network [14] and therefore the energy efficiency of this layer as well as that of the metro and the core network have to be improved to improve the overall energy efficiency of the IoT-to-cloud or IoT-to-edge processing architectures. Attention was given to the energy efficiency of different network segments [15]-[21], to the use of renewable energy in these networks to reduce $CO_2$ [22] and to different energy efficient transmission strategies [23], [24]. IoT nodes can generate large amounts of data and therefore these big data networks have to be optimized to improve their energy efficiency given data processing and networking power consumption, and these were evaluated in [25]-[28]. The exceptional amount of data generated by IoT objects is currently estimated at 2.3 trillion gigabytes of data every day [3].

Serious concerns have been raised about the cost of the energy needed to transport such huge data through the Internet so that it is accessible by anyone anywhere. The connection between the IoT objects and the Internet is facilitated by access networks. One of the most favourable access networks in terms of high bandwidth, long access distance and power consumption is passive optical networks (PON).

Energy constraints are a dominant trait of most IoT end nodes. Many of the IoT implementations use wireless for connectivity. The IoT wireless modules are well known for their hunger for energy. Therefore, processing and computation offloading to the edge of the network is a key method to save energy [3], [29]. Edge computing is proposed to assist in tackling the computational resource poverty of IoT objects. Some of previous studies and research efforts have considered

Manuscript received xxxx; revised xxxx; accepted xxxx. Date of publication xxxx; date of current version xxxx. This work was supported by the Engineering and Physical Sciences Research Council (EPSRC), INTERNET (EP/H040536/1), STAR (EP/K016873/1) and TOWS (EP/S016570/1) projects.
Zaineb T. Al-Azez, Ahmed Q. Lawey, Taisir E.H. El-Gorashi, Jaafar M.H. Elmirghani are with the School of Electronic and Electrical Engineering, University of Leeds, Leeds, LS2 9JT, U.K. (e-mail: elztaa@leeds.ac.uk; A.Q.Lawey@leeds.ac.uk; t.e.h.elgorashi@leeds.ac.uk; j.m.h.elmirghani@leeds.ac.uk).

addressing some issues such as power consumption, cost and bandwidth in IoT and PON architectures. The authors of [30] proposed a dynamic bandwidth allocation scheme for converged 5G mobile fronthall and IoT networks on TDM-PON. They proposed this scheme to address some technical issues regarding uplink bandwidth management. The work introduced in [31] proposed the implementation of monitoring and control systems in hospice environment through the use of wireless sensors and actuators modules and through the storage of the data in the cloud within a hospital. The authors of [31] proposed the integration of cloud networking with wiFi and ZigBee to realize a Wireless Hospital Digital Interface (WHDI). The authors of [32] improved cost and power consumption figures of the introduced access network by introducing a novel network architecture. The proposed architecture conquers the limitations of both long-reach PONs and mobile backhauling schemes. The enhanced architecture is based on adaptive ultra-long reach links to bypass the Metropolitan Area Network on the core side, in addition to the use of a low cost and low power consumption technology (short-range XPON, wireless) at the end user side. In order to enhance performance, open access networking models and Software Defined Networking (SDN) principles support network virtualization and efficient resource management. In [33], the authors considered some potential IoT access network technologies and examined these technologies over a range of traffic levels in term of power efficiency. The authors of [33] showed that the use of wiFi with PON backhaul, 4G wireless (LTE) access and also GPON access is the most energy efficient access architecture for different IoT traffic levels.

In this paper, we design a framework for an energy efficient edge computing platform for IoT supported by a PON. In this paper, we expand our brief initial work proposed in [34] and provide a full MILP optimization model whose details are given here for the first time. In addition, we expand the work in [34] by developing a heuristic algorithm that mimics, in real time, the behavior of the MILP model introduced.

The remainder of this paper is organized as follows. In Section II, we describe our energy efficient MILP optimization model. Section III discusses the MILP model results, Section IV presents the heuristic and Section V discusses the heuristic results. Finally, in Section VI we give our conclusions.

## II. MILP FOR ENERGY EFFICIENT PON-IOT NETWORKS

Our MILP model considers the architecture shown in Figure 1. In this architecture, the upper core network receives the aggregated processed traffic from two separate IoT networks through a PON. In our framework, each IoT network consists of four layers. IoT objects represent the first layer while relay elements represent the second layer. Relay elements aggregate the traffic from the IoT objects. The traffic from the relay elements is aggregated by a single coordinator element in the third layer. A single gateway element is hosted by the last layer in the IoT network. The coordinator traffic is aggregated by this single gateway element and uploaded to the access network (PON). In our framework, the PON access network is made up of two layers. Two ONU entities are hosted by the first layer (ONU layer) while OLT layer consists of a single OLT entity. The task of PON access network is to aggregate the traffic from the IoT network and upload it to the core network.

In our architecture, the entities in PON access network layers and the elements in the three upper layers in each IoT network are allowed to host VMs. These VMs are capable of processing the aggregated traffic. We considered different applications by modelling different types of VMs. Only one VM type is requested by each IoT object. Useful information is generated by VMs through reducing traffic at different traffic reduction percentages. This traffic reduction is done by the VMs by processing the incoming raw data.

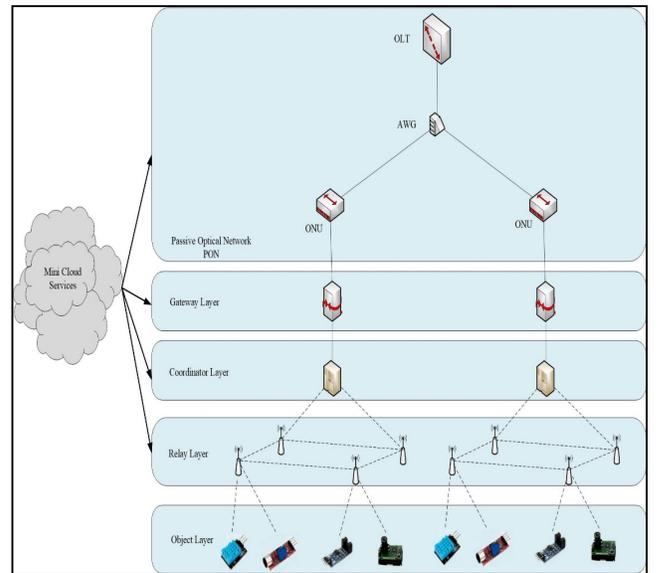

Figure 1. The PON-IoT architecture evaluated

Minimizing the total power consumption is the objective of our proposed MILP. The total power consumption consists of two basic components. Firstly, the power consumption due to traffic through all layers of the proposed architecture. Secondly, the power consumption due to processing by VMS in all possible hosting layers. The MILP power minimization is has to observe some constraints. These constraints consider cloudlet placement, the optimal placement of VMs, managing traffic direction and the traffic flow conservation for unprocessed and processed IoT traffic. Regarding the proposed MILP notations, we have used superscripts to index the variables and the parameters while we have used subscripts as indices of these variables and parameters. Table I defines the parameters used in the MILP model:

**Table I List of parameters and their definitions**

| Notation | Description |
|---|---|
| $O$ | Set of IoT objects |
| $R$ | Set of relays |
| $C$ | Set of coordinators |

| | | | |
|---|---|---|---|
| $G$ | Set of gateways | | **Table II List of variables and their definitions** |
| $ONU$ | Set of ONUs | **Notation** | **Description** |
| $OLT$ | Set of OLTs | $\lambda_{ovc}^{upt}$ | Un-processed traffic from the IoT object $o$ to the virtual machine $v$ placed at the cloudlet $c$ |
| $TN$ | Set of all IoT network nodes ($TN = O \cup R \cup C \cup G \cup ONU \cup OLT$) | | |
| $N_x$ | Set of neighbours of node $x$ ($N_x, x \in TN$) | $\lambda_{oc}^{upt}$ | Un-processed traffic from IoT object $o$ to cloudlet $c$ placed in the candidate networking element |
| $CN$ | Set of candidate nodes for the cloudlet placement ($CN = R \cup C \cup G \cup ONU \cup OLT$) | | |
| $VM$ | Set of virtual machine types | $\lambda_{ocxy}^{upt}$ | Un-processed traffic from the IoT object $o$ to cloudlet $c$ placed in the candidate networking element passing through the link between the nodes pair $(x,y)$ |
| $\lambda_{ov}^{upt}$ | Un-processed traffic from the IoT object $o$ to the virtual machine $v$, in kbps | | |
| $d_{xy}$ | Distance between the node pair $(x,y)$ in the IoT network, in meters | | |
| $\epsilon$ | Transmission amplifier power coefficient, in joule/(bit.m$^2$) | $\lambda_{xy}^{upt}$ | Un-processed traffic between the nodes pair $(x,y)$ |
| $E^{ot}$ | IoT object energy per bit for transmission, in joule/bit | $\lambda_{xy}^{pt}$ | Processed traffic between the nodes pair $(x,y)$ |
| $E^{rt}$ | Relay energy per bit for transmission, in joule/bit | $\lambda_{cl}^{pt}$ | Processed traffic from cloudlet $c$ placed in the candidate networking element to the OLT $l$ |
| $E^{rr}$ | Relay energy per bit for receiving, in joule/bit | | |
| $E^{ct}$ | Coordinator energy per bit for transmission, in joule/bit | $\lambda_{clxy}^{pt}$ | Processed traffic from cloudlet $c$ placed in the candidate networking element to the OLT $l$ passing through the link between the nodes pair $(x,y)$ |
| $E^{cr}$ | Coordinator energy per bit for receiving, in joule/bit | | |
| $E^{gr}$ | Gateway energy per bit for receiving, in joule/bit | | |
| $E^{gt}$ | Gateway energy per bit for transmitting, in joule/bit | $I_{vc}$ | $I_{vc} = 1$ if the virtual machine $v$ is placed in the cloudlet $c$, otherwise $I_{vc} = 0$ |
| $E^u$ | ONU energy per bit, in joule/bit | $H_c$ | $H_c = 1$ if a cloudlet $c$ is built at the candidate networking element, otherwise $H_c = 0$ |
| $E^l$ | OLT energy per bit, in joule/bit | | |
| $W_{vc}$ | Normalized workload of the virtual machine $v$ in cloudlet $c$ | $TW_c$ | Total normalized workload of the cloudlet $c$ built at candidate networking element |
| $RMP$ | Maximum processing power consumption at relay elements | $PC^{rp}$ | Total processing induced power consumption of the relays |
| $CMP$ | Maximum processing power consumption at coordinator elements | $PC^{cp}$ | Total processing induced power consumption of the coordinators |
| $GMP$ | Maximum processing power consumption at gateway elements | $PC^{gp}$ | Total processing induced power consumption of the gateways |
| $UMP$ | Maximum processing power consumption at ONU entities | $PC^{up}$ | Total processing induced power consumption of the ONUs |
| $LMP$ | Maximum processing power consumption at OLT entities | $PC^{lp}$ | Total processing induced power consumption of the OLTs |
| $\gamma, \beta$ | Large enough numbers | | |
| $F$ | Traffic reduction factor | $PC^{otr}$ | Total traffic induced power consumption of the IoT objects |
| $A$ | Networking elements scaling factor | | |

| Symbol | Description |
|---|---|
| $PC^{rtr}$ | Total traffic induced power consumption of the relays |
| $PC^{ctr}$ | Total traffic induced power consumption of the coordinators |
| $PC^{gtr}$ | Total traffic induced power consumption of the gateways |
| $PC^{utr}$ | Total traffic induced power consumption of the ONUs |
| $PC^{ltr}$ | Total traffic induced power consumption of the OLTs |

The total IoT processing induced power consumption is composed of:

1) The processing induced power consumption of each relay:
$$PC^{rp} = TW_c \cdot RMP \quad (1)$$
$$\forall c \in R$$

2) The processing induced power consumption of each coordinator:
$$PC^{cp} = TW_c \cdot CMP \quad (2)$$
$$\forall c \in C$$

3) The processing induced power consumption of each gateway:
$$PC^{gp} = TW_c \cdot GMP \quad (3)$$
$$\forall c \in G$$

4) The processing induced power consumption of each ONU:
$$PC^{up} = TW_c \cdot UMP \quad (4)$$
$$\forall c \in ONU$$

5) The processing induced power consumption of the OLT:
$$PC^{lp} = TW_c \cdot LMP \quad (5)$$
$$\forall c \in OLT$$

The processing induced power consumption of all processing elements in our proposed network (relays, coordinators, gateways, ONUs and OLT) are evaluated in equations (1) to (5). The processing induced power consumption of each element is a function of its CPU maximum power and total normalized workload utilization of the cloudlet placed in the element.

The total IoT traffic induced power consumption is composed of:

1) The traffic induced power consumption of each IoT object :
$$PC^{otr} = \sum_{y \in R} \lambda_{xy}^{upt} \cdot (E^{ot} + \epsilon \cdot d_{xy}^2) \quad (6)$$
$$\forall x \in O$$

2) The traffic induced power consumption of each relay:
$$PC^{rtr} = \sum_{y \in R \cup C: y \neq x} (\lambda_{xy}^{upt} + \lambda_{xy}^{pt}) \cdot (E^{rt} + \epsilon \cdot d_{xy}^2) + \sum_{y \in O \cup R: y \neq x} (\lambda_{yx}^{upt} + \lambda_{yx}^{pt}) \cdot E^{rr} \quad (7)$$
$$\forall x \in R$$

3) The traffic induced power consumption of each coordinator:
$$PC^{ctr} = \sum_{y \in G} (\lambda_{xy}^{upt} + \lambda_{xy}^{pt}) \cdot (E^{ct} + \epsilon \cdot d_{xy}^2) + \sum_{y \in R} (\lambda_{yx}^{upt} + \lambda_{yx}^{pt}) \cdot E^{cr} \quad (8)$$
$$\forall x \in C$$

4) The traffic induced power consumption of each gateway:
$$PC^{gtr} = \sum_{y \in ONU} (\lambda_{xy}^{upt} + \lambda_{xy}^{pt}) \cdot E^{gt} + \sum_{y \in C} (\lambda_{yx}^{upt} + \lambda_{yx}^{pt}) \cdot E^{gr} \quad (9)$$
$$\forall x \in G$$

5) The traffic induced power consumption of each ONU:
$$PC^{utr} = \sum_{y \in OLT} (\lambda_{xy}^{upt} + \lambda_{xy}^{pt}) \cdot E^u + \sum_{y \in G} (\lambda_{yx}^{upt} + \lambda_{yx}^{pt}) \cdot E^u \quad (10)$$
$$\forall x \in ONU$$

6) The traffic induced power consumption of the OLT:
$$PC^{ltr} = \sum_{y \in ONU} (\lambda_{yx}^{upt} + \lambda_{yx}^{pt}) \cdot E^l \quad (11)$$
$$\forall x \in OLT$$

Traffic induced power consumption components of our proposed network are represented by equations (6) to (11). The general structure of these equations is based on radio energy dissipation equation (Friis free-space equation) used in [35]. These equations are comprised of two basic parts the sending part and receiving part. Both parts are based on bit rate times the propagation energy per bit. Equation (6) represents the traffic induced power consumption of the IoT objects. This equation considers the sending traffic only because the traffic received by the IoT objects is considered in this model as signalling messages with small data size that can be ignored. On the other hand, equation (11) considers only the receiving traffic induced power consumption of OLT as the OLT layer is the highest layer in the model.

**Objective: Minimize**

$$\sum_{c \in R} PC^{rp} + \sum_{c \in C} PC^{cp} + \sum_{c \in G} PC^{gp}$$
$$+ \sum_{\forall c \in ONU} PC^{up}$$
$$+ \sum_{\forall c \in OLT} PC^{lp} + \sum_{x \in O} PC^{otr}$$
$$+ A$$
$$\cdot \left( \sum_{x \in R} PC^{rtr} + \sum_{x \in C} PC^{ctr} \right.$$
$$+ \sum_{x \in ONU} PC^{utr}$$
$$\left. + \sum_{x \in OLT} PC^{ltr} \right)$$
$$+ \sum_{x \in G} PC^{gtr} \quad (12)$$

The model objective is to minimize the PON and IoT network power consumption due to traffic processing and aggregation as presented in equation (12). The scaling factor A is introduced to examine the case where the traffic induced power consumption in the networking elements is comparable to their processing induced power consumption.

**Subject to:**
1) IoT network un-processed traffic constraints

$$\sum_{\forall c \in CN} \lambda_{ovc}^{upt} = \lambda_{ov}^{upt} \quad (13)$$
$$\forall o \in O, \forall v \in VM$$

$$\lambda_{oc}^{upt} = \sum_{\forall v \in VM} \lambda_{ovc}^{upt} \quad (14)$$
$$\forall o \in O, \forall c \in CN$$

$$\sum_{y \in N_x} \lambda_{ocxy}^{upt} - \sum_{y \in N_x} \lambda_{ocyx}^{upt}$$
$$= \begin{cases} \lambda_{oc}^{upt} & \text{if } x = o \\ -\lambda_{oc}^{upt} & \text{if } x = c \\ 0 & \text{otherwise} \end{cases} \quad (15)$$
$$\forall o \in O, \forall c \in CN, \forall x \in TN$$

$$\lambda_{xy}^{upt} = \sum_{c \in CN} \sum_{o \in O} \lambda_{ocxy}^{upt} \quad (16)$$
$$\forall x \in TN, \forall y \in N_x$$

Constraint (13) distributes the unprocessed traffic from IoT objects ($o$) over a number of VM ($v$) instances that are hosted in different mini cloudlets ($c$). It ensures that the total un-processed traffic flows from the IoT object ($o$) to all VM ($v$) instances in different mini cloudlets ($c$) equals to the traffic between that object ($o$) and the VM ($v$). Constraint (14) calculates the traffic flowing from IoT objects to each networking element. It ensures that the total un-processed traffic from the IoT object $o$ to all the virtual machines $v$ placed in cloudlet $c$ is equal to the un-processed traffic from the object $o$ to cloudlet $c$ placed in candidate networking element. Constraint (15) represents the flow conservation for the un-processed traffic from the IoT object $o$ to cloudlet $c$ located in candidate networking element. It ensures that the total un-processed outgoing traffic is equal to the total un-processed incoming traffic for each IoT node except for the source and the destination. Constraint (16) represents the total unprocessed traffic between any IoT node pair ($x$,$y$).

2) IoT network processed traffic constraints

$$\sum_{\forall l \in OLT: c \notin OLT} \lambda_{cl}^{pt} = F \cdot \sum_{\forall o \in O} \lambda_{oc}^{upt} \quad (17)$$
$$\forall c \in CN$$

$$\sum_{y \in N_x \cap CN} \lambda_{clxy}^{pt} - \sum_{y \in N_x \cap CN} \lambda_{clyx}^{pt}$$
$$= \begin{cases} \lambda_{cl}^{pt} & \text{if } x = c \\ -\lambda_{cl}^{pt} & \text{if } x = l \\ 0 & \text{otherwise} \end{cases} \quad (18)$$
$$\forall c \in CN, \forall l \in OLT, \forall x \in CN: c \neq l$$

$$\lambda_{xy}^{pt} = \sum_{c \in CN} \sum_{l \in OLT: c \neq l} \lambda_{clxy}^{pt} \quad (19)$$
$$\forall x \in CN, \forall y \in N_x \cap CN$$

Constraint (17) calculates the reduced traffic flowing from the candidate networking element hosted in cloudlet $c$ to the OLT $l$. Constraint (18) represents the flow conservation for the processed traffic from the candidate networking element hosted cloudlet $c$ to the OLT $l$. It ensures that the total processed outgoing traffic is equal to the total processed incoming traffic for each IoT and PON node except for the source and the destination. Constraint (19) represents the total processed traffic between any IoT and PON node pair ($x$,$y$).

3) Virtual machine placement and workload constraints

$$\sum_{o \in O} \lambda_{ovc}^{upt} \geq I_{vc} \quad (20)$$
$$\forall v \in VM, \forall c \in CN$$

$$\sum_{o \in O} \lambda_{ovc}^{upt} \leq \beta \cdot I_{vc} \quad (21)$$
$$\forall v \in VM, \forall c \in CN$$

$$\sum_{v \in VM} I_{vc} \geq H_c \quad (22)$$
$$\forall c \in CN$$

$$\sum_{v \in VM} I_{vc} \leq \gamma \cdot H_c \quad (23)$$
$$\forall c \in CN$$

$$TW_c = \sum_{v \in VM} W_{vc} \cdot I_{vc} \quad (24)$$

$$\forall c \in CN$$

Constraints (20) and (21) place the virtual machine $v$ in the cloudlet $c$ if the cloudlet $c$ is serving some IoT object requests for this virtual machine. $\beta$ is a large enough number with units of bps to ensure that $I_{vc}= 1$ when $\sum_{o \in O} \lambda_{ovc}^{upt}$ is greater than zero, otherwise $I_{vc}= 0$. Constraints (22) and (23) build a cloudlet $c$ in the candidate networking element if this networking element is chosen to host at least one virtual machine $v$, where $\gamma$ is a large enough unitless number to ensure that $H_c= 1$ if $\sum_{v \in VM} I_{vc}$ is greater than zero, otherwise $H_c=0$. Constraint (24) calculates the total normalized workload of each built cloudlet $c$.

## III. MILP EVALUATION AND RESULTS

In our evaluation, two IoT networks were considered, supported a PON network. The scenario has in each IoT network: 50 IoT objects, 25 relays, one gateway and a single coordinator. The PON OLT supports two ONUs and each IoT network is connected to an ONU. Figure 2 shows a 30 m × 30 m area which contains the components of each IoT network, namely the IoT objects, relays and coordinator, with 100m separating the gateway from the coordinator. The IoT objects are distributed in the 30 m × 30 m space randomly and uniformly, while the relays are separated by 6 m in a deterministic and uniform fashion as shown in Figure 2. Communication in the IoT network uses the Zigbee protocol which supports the IoT devices. A Gigabit Ethernet link is used to connect the gateway to the ONU. The ONU to OLT fibre link is part of the PON architecture. In our study we only consider the uplink direction as most of the IoT traffic is carried in this direction. As such, we also consider a setup where traffic does not pass from an IoT network to another IoT network through the OLT. We consider in our model the power consumption in the PON modules (ONUs and OLT) due to the traffic flowing in the network. We also consider the power consumption of the IoT network components attributable to transmitters, receivers and power amplifiers which compensate for the propagation losses incorporated in our models [36].

Table III summarizes the parameters used in the model. In terms of power consumption, two parts are considered for each network element in the proposed network; namely the communication and processing parts. The specifications of communication part used in objects, relays and coordinator are based on [37] while we used Cisco 910 industrial router [38] for the communication part of the gateway. In addition we used FTE7502 EPON ONU [39] and FSU7100 EPON OLT [40] as the ONU and OLT elements in the proposed network. The relays, coordinator, gateway, ONU and OLT elements are equipped with Intel Atom Z510 CPU [41] used for processing. We have considered a range of traffic reduction percentages after processing in order to investigate different impacts of processing applications.

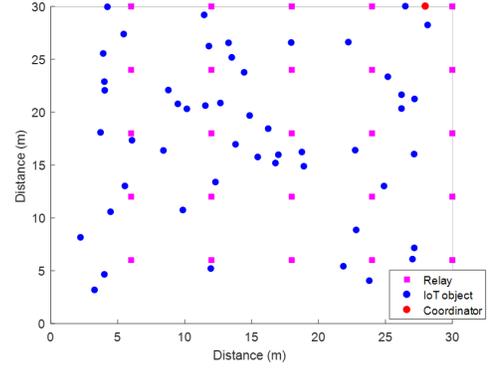

Figure 2 IoT objects and the distribution of relays

**Table III List of input parameters**

| Parameter Name | Value |
|---|---|
| Traffic sent from IoT object to a VM type ($\lambda_{ov}^{upt}$) | 5 kbps [42] |
| CPU maximum power consumption (RMP, CMP, GMP) | 4.64 W [41] |
| Number of CPUs used in a relay, coordinator, gateway, ONU and OLT. | 1, 2, 4, 4, 10 |
| IoT object, relay and coordinator transmitting energy per bit ($E^{ot}, E^{rt}, E^{ct}$) | 50 nJ/bit [37] |
| Relay and coordinator receiving energy per bit ($E^{or}, E^{rr}, E^{cr}$) | 50 nJ/bit [37] |
| Gateway receiving energy per bit ($E^{gr}$) | 60 µJ/bit [38] |
| Gateway sending energy per bit ($E^{gt}$) | 15 nJ/bit [38] |
| ONU energy per bit ($E^u$) | 7.5 nJ/bit [40] |
| OLT energy per bit ($E^l$) | 225.6 pJ/bit [40] |
| Transmission amplifier power coefficient ($\epsilon$) | 255 pJ/(bit.$m^2$) [37] |
| VM type 1 normalized workload in relay, coordinator, gateway, ONU and OLT elements ($W_{1c}$) | 0.1, 0.05, 0.025, 0.025, 0.01 [43] |
| VM type 2 normalized workload in relay, coordinator, gateway, ONU and OLT elements ($W_{2c}$) | 0.2, 0.1, 0.05, 0.05, 0.02 [43] |
| VM type 3 normalized workload in relay, coordinator, gateway, ONU and OLT elements ($W_{3c}$) | 0.3, 0.15, 0.075, 0.075, 0.03 [43] |
| VM type 4 normalized workload in relay, coordinator, gateway, ONU and OLT elements ($W_{4c}$) | 0.4, 0.2, 0.1, 0.1, 0.04 [43] |
| Traffic reduction percentage (F) | {10, 30, 50, 70, 90}% |
| Distance between node pair (x, y) in the IoT network, in meters ($d_{xy}$) | Within 30 m × 30 m [44] |
| $\gamma, \beta, A$ | 50, 10000000 bps, 5 |

In our evaluation three scenarios were considered. In the first scenario four VM types were considered characterized by heterogeneous VM CPU demands that range from 10% CPU utilization to 40% CPU utilization. Homogeneous, 40%, CPU requirements were considered in the second scenario which has four VM types. In the third scenario, a setup similar to that of Scenario 2 was considered where there are four VM types, all homogeneous, and require 40% of the CPU. In the third scenario however there is access to the OLT which has a CPU. The OLT CPU has a lower energy efficiency, (requiring 9.28W), but has similar processing capabilities. These scenarios make it possible to evaluate our framework at different equipment energy efficiency levels and different CPU demands. The power consumption due to processing, the power consumption due to traffic and the total power consumption are shown in Figures 3, 4 and 5. The optimum placement of VMs in the three scenarios are shown in Figures 6, 7 and 8. In Scenario 1, heterogeneous VMs are considered. This scenario is able to place some of the VMs (at 10% traffic reduction) in the OLT as shown in Figure 6. It results in the lowest processing power consumption among the three scenarios. The low power consumption in this scenario is due to the reduction in the total number of VMs needed. Placing VMs in the OLT, whenever they fit, allows the VMs to be shared thus reducing the total number of VMs needed and hence reducing the power consumption. Note that VM placement in any other layer leads to VM duplication as traffic is not allowed to pass between the IoT networks due to the limited downlink capacity as discussed earlier.

In Scenario 2 the OLT hosts more VMs (10% traffic reduction case, Figure 7) due to the homogeneous CPU utilization in this case. Despite this, Scenario 2 however has higher CPU power consumption in comparison with Scenario 1. This is attributed to the fact that the collection of VMs in Scenario 2 consume higher power at the 10% traffic reduction compared to Scenario 1 as shown in Figure 3.

Examining Figure 3 shows that Scenario 3 has the highest CPU power consumption at the low (10%) traffic reduction percentage. Observe that the OLT has an energy inefficient CPU. This results in the VMs being placed at the lower layers as can be seen in Figure 8 (10% case).

It should be noted that at high traffic reduction ratios (50% - 90%), the VMs are placed in all scenarios in the relay layer in both IoT networks as shown in Figures 6, 7 and 8. This choice results in the minimum power consumption due to traffic as the higher layers are not accessed.

Scenario 1 maintains the lowest CPU power consumption compared to the other two scenarios as it considers heterogeneous VMs. The other two scenarios have comparable CPU power consumption (in the 30% to 90% range in Figure 3) since both scenarios (Scenarios 2 and 3) serve VMs that have similar CPU utilization using the relay elements.

Figure 4 shows that the network power consumption is progressively reduced as the traffic reduction ratio increases. This is attributed to the smaller traffic volume which induces lower power consumption as the traffic reduction ratio increases. In this case more segments of the network carry the smaller, extracted knowledge, instead of the raw unprocessed traffic.

In Figure 4 also note that Scenario 3 has the lowest power consumption attributable to traffic at the 10% traffic reduction ratio. Scenario 3 is able to place more VMs in the coordinator layer compared to the other scenarios, see Figures 6, 7 and 8 and the 10% traffic reduction case, hence more the knowledge-bearing lower-volume traffic passes to the upper layers. It has to be noted however that this reduction in network power consumption in Scenario 3 is overwhelmed by the increase in CPU power consumption at low reduction percentage, this leading to higher overall power consumption in Scenario 3 compared to the other two scenarios, see Figure 5 at the 10% traffic reduction case. In terms of traffic induced power consumption (at low traffic reduction ratio, ie 10%, see Figure 4), the next best is Scenario 1 which places some VMs in the lower network layers (10% case in Figure 6). Scenario 2 places the VMs in the OLT (10% case in Figure 7) which leads to the highest power consumption attributable to traffic (the 10% traffic reduction case in Figure 4).

Furthermore, Scenario 1 results in the placement of more cloudlets in the relay layer compared to the other scenarios (70% traffic reduction case in Figures 6, 7 and 8). This leads to a slightly higher power consumption attributable to traffic, see the 70% case in Figure 4. It can also be noted that comparable power consumption due to traffic is observed in Figure 4 for all scenarios at 30%, 50% and 90% traffic reduction ratios.

This is attributable to the VMs distribution which is similar in Figures 6, 7 and 8 in all these cases. This similar distribution results from the high traffic reduction ratios in these cases which lead to the placement of the VMs in the relay layer, ie the layer closest to the IoT objects, to capitalise on this reduction in traffic. Figure 5 shows that Scenario 1 is the most energy efficient scenario overall. It has the lowest power consumption attributable to processing, and this more than compensates for its higher traffic induced power consumption. As a result, Scenario 1 has 17% and 19% total power consumption savings compared to Scenarios 2 and 3 respectively.

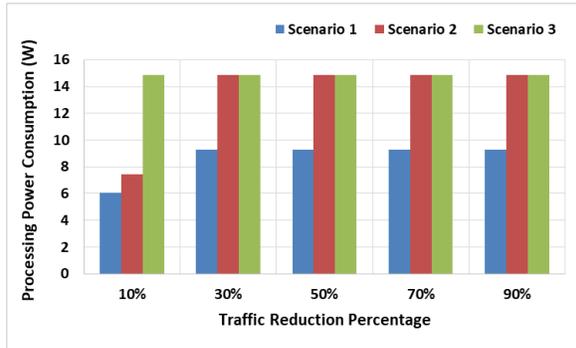

**Figure** 3 Processing power consumption of the three scenarios

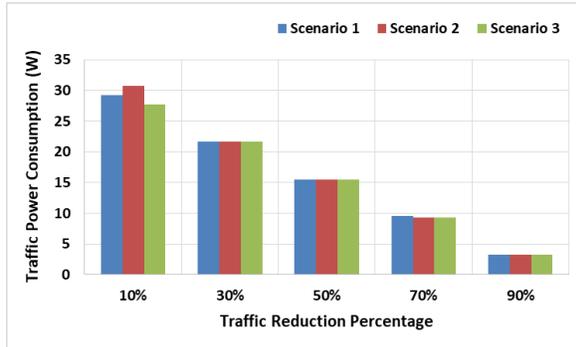

**Figure** 4 Traffic power consumption of the three scenarios

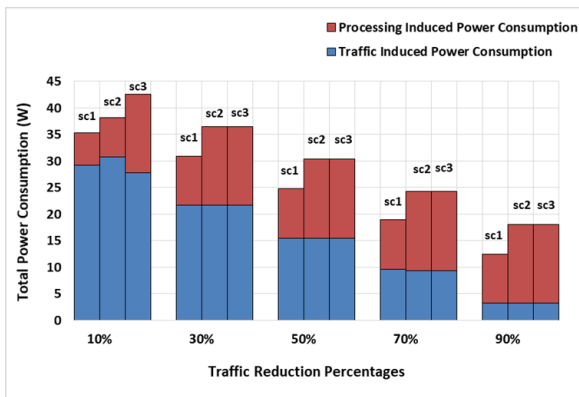

**Figure** 5 Total power consumption of the three scenarios

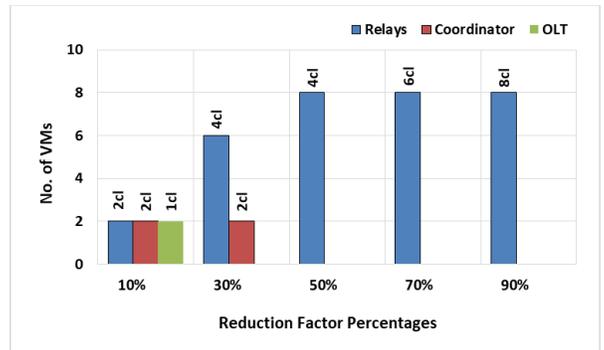

**Figure** 6 VMs placement in different cloudlets (cl) in Scenario 1

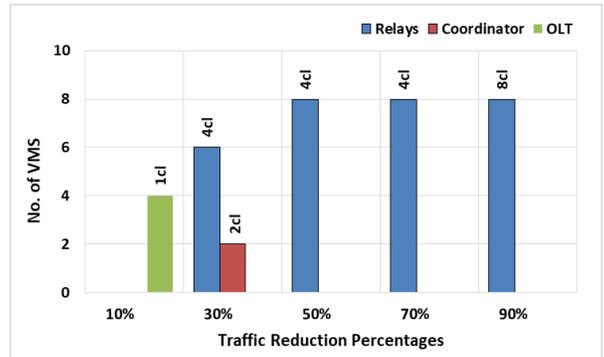

**Figure** 7 VMs placement in different cloudlets (cl) in Scenario 2

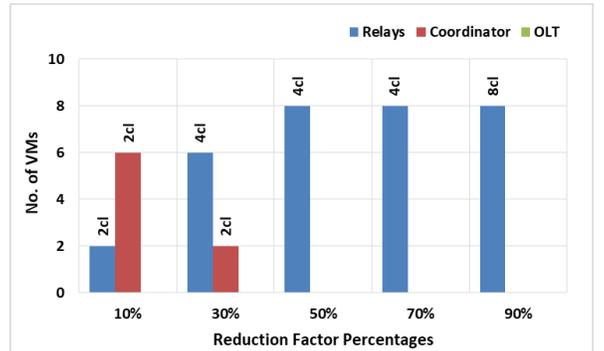

**Figure** 8 VMs placement in different cloudlets (cl) in Scenario 3

## IV. EEPIV HEURISTIC

This section validates the MILP model results by presenting the Energy Efficient PON supported IoT Virtualization (EEPIV) heuristic that mimics the MILP model behavior. The pseudo code of the EEPIV heuristic is presented in Figure 9. The heuristic shown in Figure 9 covers all the scenarios of our MILP model as implementing these scenarios relies on changing the input parameters not the constraints that the model is subject to.

The heuristic calculates the total power consumption (TPC) of the network according the optimum place and the number of mini cloudlets that serve the IoT objects through the hosted VMs. Serving IoT objects by VMs is subject to the limited capabilities of the serving host VM in each cloudlet as below:

i. There should be sufficient processor capacity in each candidate cloudlet to accommodate the hosted VM workload.

ii. The intended VM $v$ that is requested by IoT object $o$ in each network should not have been hosted by any other cloudlet in this network before.

If all the serving constraints above are met, then the heuristic hosts the intended VM in the candidate cloudlet to satisfy the IoT object request and sets the binary indicator $F_{cv}$ accordingly. The total workload of each hosted cloudlet in the candidate place is calculated depending on the binary indicator $F_{cv}$.

Since the processing induced power consumption of each processing element is a function of the total workload of the cloudlet, the heuristic calculates the processing induced power consumption of all the processing elements in the proposed network (relays, coordinators, gateways, ONUs and OLT) as shown in steps 11 to 35 in Figure 9. The end-to-end traffic generated by the IoT objects' requests is next calculated by the heuristic. The traffic passes through two stages: the first stage flows from the generator (IoT object) to the destined VM in the hosting cloudlet which is represented by $\lambda_{oc}^{upt}$ (unprocessed traffic). The second stage comes after the processing stage. In this stage, the processed traffic $\lambda_{cl}^{pt}$ (reduced traffic) flows from the cloudlet to the last layer in the network which is represented in our proposed network by the OLT layer. The intermediate traffic between each node pair in the network is calculated by the heuristic model based on the end to end traffic. The heuristic routes the traffic through these intermediate nodes from the source to the destination using a minimum hop algorithm to reduce the traffic induced power consumption. Finally, the heuristic calculates the total power consumption $TPC$ by summing all the processing and traffic induced power consumption of all nodes.

---

**Inputs:** $VM = \{1 \ldots NVM\}$

$CN = \{1 \ldots NCN\}$

$O = \{1 \ldots NO\}$

$R = \{1 \ldots NR\}$

$C = \{1 \ldots NC\}$

$G = \{1 \ldots NG\}$

$ONU = \{1 \ldots NONU\}$

$OLT = \{1 \ldots NOLT\}$

**Output:** No. of Served Objects

        Total Power Consumption (TPC)

1.   **For** each candidate cloudlet that can host a required VM $c \in CN$ **Do**

2.   **For** each Virtual Machine required by an object $v \in VM$ **Do**

3.     **If** $U_{ov} > 0$ **Then**

4.       **If** all serving constraints are met **Then**

5.         $F_{cv}(c, v) = 1$

6.         Calculate the workload of the hosting cloudlet $TCWc$

        without considering the number of served IoT objects

7.       **End If**

8.     **End If**

9.   **End For**

10.  **End For**

11.  **For** Each relay ($r \in R$) **Do**

12.     **If** the hosting cloudlet is placed in relay layer R $c \in CN$ **Do**

13.       Calculate R_PPC

14.     **End If**

15.  **End For**

16.  **For** Each coordinator ($c \in C$) **Do**

17.     **If** the hosting cloudlet is placed in coordinator layer C $c \in CN$ **Do**

18.       Calculate C_PPC

19.     **End If**

20.  **End For**

21.  **For** each gateway ($g \in G$) **Do**

22.     **If** the hosting cloudlet is placed in gateway layer $c \in CN$ **Do**

23.       Calculate G_PPC

24.     **End If**

25.  **End For**

26.  **For** each ONU ($u \in ONU$)

27.     **If** the hosting cloudlet is placed in ONU layer $c \in CN$ **Do**

28.       Calculate $PC^{up}$

29.     **End If**

30.  **End For**

31.  **For** each OLT ($l \in OLT$)

32.     **If** the hosting cloudlet is placed in OLT layer $c \in CN$ **Do**

33.       Calculate $PC^{up}$

| | |
|---|---|
| 34.        **End If** | 62.        Calculate $ONU\_tr$ based on minimum hop path between node |
| 35.   **End For** |         pair (x,y) |
| 36.   **For** each IoT object served by a cloudlet $o \in O$ **Do** | 63.   **End For** |
| 37.      **For** each hosting cloudlet $c \in CN$ **Do** | 64.   Calculate total power consumption |

36. **For** each IoT object served by a cloudlet $o \in O$ **Do**
37.    **For** each hosting cloudlet $c \in CN$ **Do**
38.      Calculate end to end traffic that flows from each object to the
         cloudlet that serves this object
39.    **End For**
40.  **End For**
41. **For** Each hosting cloudlet $c \in CN$ **Do**
42.    **For** Each OLT ($l \in OLT$) **Do**
43.      Calculate the end to end reduced traffic from the cloudlet to
         the OLT
44.    **End For**
45.  **End For**
46. **For** each IoT object $o \in O$ **Do**
47.    Calculate TO_tr
48. **End For**
49. **For** each relay ($r \in R$)
50.    Calculate $TR\_tr$ based on minimum hop path between node
       pair (x,y)
51. **End For**
52. **For** each coordinator ($c \in C$)
53.    Calculate $TC\_tr$ based on minimum hop path between node
       pair (x,y)
54. **End For**
55. **For** each gateway ($g \in G$)
56.    Calculate $TG\_tr$ based on minimum hop path between node
       pair (x,y)
57. **End For**
58. **For** each ONU ($u \in ONU$)
59.    Calculate $ONU\_tr$ based on minimum hop path between node
       pair (x,y)
60. **End For**
61. **For** each OLT ($l \in OLT$)
62.    Calculate $ONU\_tr$ based on minimum hop path between node
       pair (x,y)
63. **End For**
64. Calculate total power consumption

$$TPC = \sum_{r \in R} R\_PPC + \sum_{c \in C} C\_PPC + \sum_{g \in G} G\_PPC + \sum_{u \in ONU} ONU\_PPC + \sum_{o \in O} TO\_tr + \sum_{r \in R} TR\_tr + \sum_{c \in C} TC\_tr + \sum_{g \in G} TG\_tr + \sum_{u \in ONU} ONU\_tr + \sum_{l \in OLT} OLT\_tr$$

**Figure** 9 pseudo code of EEPIV heuristic

## V. EEPIV HEURISTIC RESULTS

We used the same inputs in Table III for the heuristic. The heuristic results show close agreement with the MILP results comparing Figure 10 with Figure 5.

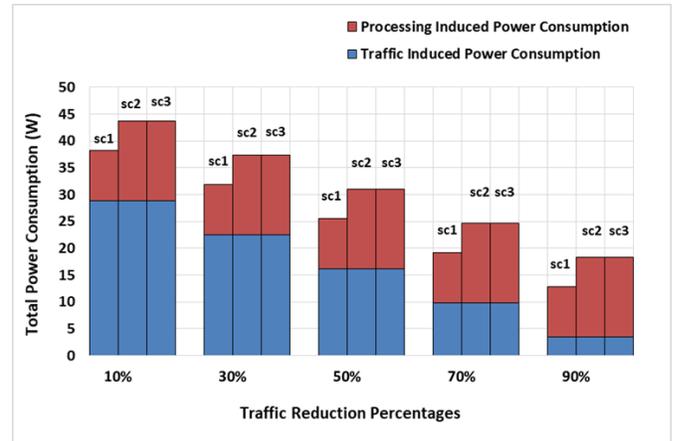

**Figure** 10 Total power consumption of the three scenarios in the heuristic

Scenarios 1 and 2 result in lower processing induced power consumption in MILP than in the heuristic at low reduction percentages (10%, Figures 3 and 11). This results from placing/using more VM copies in the heuristic (8 VMs) than in the MILP as shown in Figures 6 and 7.

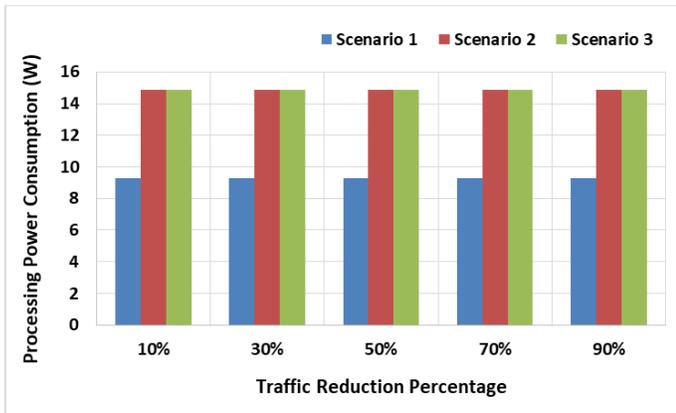

**Figure** 11 Processing power consumption of the three scenarios in the heuristic

Scenarios 1 and 2 in heuristic result in lower traffic induced power consumption than in MILP at traffic reduction percentages of 10% (Figures 4 and 12). This results from the MILP placing the serving VMs in cloudlets at higher layers (Figures 6 and 7) while all the cloudlets in the heuristic are distributed throughout the lower layer (relay layer). Placing cloudlets in higher layers results in sending more unprocessed traffic (unreduced traffic) to higher layers which in turn results in higher traffic induced power consumption. However, all the scenarios in the heuristic consume higher traffic induced power than in the MILP for the rest of the reduction percentage values as a result of the different distribution of the cloudlets in the proposed network. Since for each cloudlet, the heuristic attempts to place it in the first network element that can accommodate this cloudlet, the heuristic placed all the cloudlets in the relay layer without consideration of the closeness of the cloudlet to the IoT objects. On other side, the MILP places the cloudlets in an optimum way to minimize the traffic and processing induced power consumed by all elements of the proposed network.

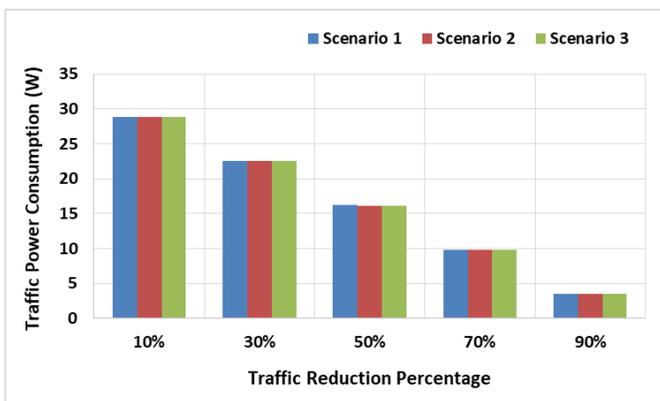

**Figure** 12 Traffic power consumption of the three scenarios in the heuristic

The number of cloudlets placed using the heuristic (2 cloudlets in Scenario 1, and 4 cloudlets in Scenarios 2 and 3 for all traffic reduction percentage values with 8 VMs in all scenarios) is less than in the MILP as one of main processes in the heuristic is bin packing where VMs must be packed into a finite number of bins (cloudlets) in a way that minimizes the number of bins used and hence the processing power consumption.

VI. CONCLUSIONS

We have developed a MILP model to optimize the placement of processing tasks in IoT networks supported by a PON infrastructure in order to minimize the overall power consumption. The model optimizes the number and location of cloudlets that are created to host the VMs where the tasks are processed. The total power consumption is made up of the power consumption due to processing and the power consumption due to traffic routing. The IoT data is reduced in volume after processing which leads to lower traffic induced power consumption. This traffic reduction after processing further influences the placement of VMs at different traffic reduction ratios. If the traffic reduction ratio is small, our results indicate that the best location to place the processing, ie the VMs is a location that can be shared by all or the majority of the IoT devices. In our architecture, this location is the OLT, which in the case consolidates the processing offered by the VMs. At the other end, when the traffic reduction ratio is high, ie if the traffic after processing is much smaller than the traffic before processing, our results show that the VMs (ie the processing) has to be placed as close to the IoT objects as possible. In our architecture this is the relay layer. Our results indicate that a power saving of up to 19% can be obtained by placing the processing (VMs) at the appropriate locations in the given setup. We have developed a heuristic for the placement of VMs and routing of information. It serves two purposes. Firstly, the heuristic is simple and therefore enables fast operation. Secondly, it acts to verify the MILP where we observed very close agreement between the heuristic and the MILP. In particular, Scenario 1 has resulted in power savings when using the heuristic of 17% (MILP 17%) and 17% (MILP 19%) compared to Scenarios 2 and 3.

ACKNOWLEDGMENTS

We would like to acknowledge funding from the Engineering and Physical Sciences Research Council (EPSRC) for the INTERNET (EP/H040536/1), STAR (EP/K016873/1) and TOWS (EP/S016570/1) projects. The first author Dr. Zaineb Al-Azez would like to thank Dr. Ahmed Al-Quzweeni for many helpful discussions and the Higher Committee for Education Development in Iraq (HCED) for funding her PhD scholarship. All data are provided in full in the results section of this paper.

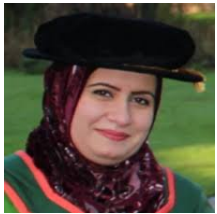
**Zaineb T. Al-Azez** received the B.Sc. and M.Sc. degrees in computer engineering from Nahrain University, Baghdad, Iraq, in 2005 and 2008, respectively, and the Ph.D. degree in communication networks from the School of Electronic and Electrical Engineering, University of Leeds, U.K., in 2018. From 2009 to 2014, she was an Assistant Lecturer with Al-Mansour University College, Baghdad. Her current research interests include energy efficiency in the Internet of Things networks, cloud computing, and virtualization.

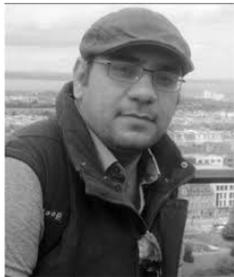
**Ahmed Q. Lawey** received the B.Sc. and M.Sc. degrees (Hons.) in computer engineering from Nahrain University, Iraq, in 2002 and 2005, respectively, and the Ph.D. degree in communication networks from the University of Leeds, U.K., in 2015. From 2005 to 2010, he was a Core Network Engineer with ZTE Corporation for Telecommunication, Iraq Branch. He is currently a Lecturer in communication networks with the School of Electronic and Electrical Engineering, University of Leeds. His current research interests include energy efficiency in optical and wireless networks, big data, cloud computing, and the Internet of Things.

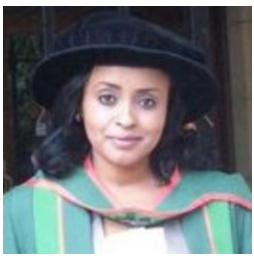
**Taisir EL-Gorashi** received the B.S. degree (first-class Hons.) in electrical and electronic engineering from the University of Khartoum, Khartoum, Sudan, in 2004, the M.Sc. degree (with distinction) in photonic and communication systems from the University of Wales, Swansea, UK, in 2005, and the PhD degree in optical networking from the University of Leeds, Leeds, UK, in 2010. She is currently a Lecturer in optical networks in the School of Electrical and Electronic Engineering, University of Leeds. Previously, she held a Postdoctoral Research post at the University of Leeds (2010– 2014), where she focused on the energy efficiency of optical networks investigating the use of renewable energy in core networks, green IP over WDM networks with datacenters, energy efficient physical topology design, energy efficiency of content distribution networks, distributed cloud computing, network virtualization and Big Data. In 2012, she was a BT Research Fellow, where she developed energy efficient hybrid wireless-optical broadband access networks and explored the dynamics of TV viewing behavior and program popularity. The energy efficiency techniques developed during her postdoctoral research contributed 3 out of the 8 carefully chosen core network energy efficiency improvement measures recommended by the GreenTouch consortium for every operator network worldwide. Her work led to several invited talks at GreenTouch, Bell Labs, Optical Network Design and Modelling conference, Optical Fiber Communications conference, International Conference on Computer Communications, EU Future Internet Assembly, IEEE Sustainable ICT Summit and IEEE 5G World Forum and collaboration with Nokia and Huawei.

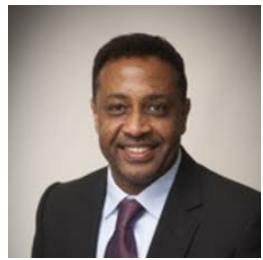
**Jaafar M. H. Elmirghani** is the Director of the Institute of Communication and Power Networks within the School of Electronic and Electrical Engineering, University of Leeds, UK. He joined Leeds in 2007 and prior to that (2000–2007) as chair in optical communications at the University of Wales Swansea he founded, developed and directed the Institute of Advanced Telecommunications and the Technium Digital (TD), a technology incubator/spin-off hub. He has provided outstanding leadership in a number of large research projects at the IAT and TD. He received the Ph.D. in the synchronization of optical systems and optical receiver design from the University of Huddersfield UK in 1994 and the DSc in Communication Systems and Networks from University of Leeds, UK, in 2012. He has co-authored Photonic Switching Technology: Systems and Networks, (Wiley) and has published over 500 papers. He has research interests in optical systems and networks. Prof. Elmirghani is Fellow of the IET, Fellow of the Institute of Physics and Senior Member of IEEE. He was Chairman of IEEE Comsoc Transmission Access and Optical Systems technical committee and was Chairman of IEEE Comsoc Signal Processing and Communications Electronics technical committee, and an editor of IEEE Communications Magazine. He was founding Chair of the Advanced Signal Processing for Communication Symposium which started at IEEE GLOBECOM'99 and has continued since at every ICC and GLOBECOM. Prof. Elmirghani was also founding Chair of the first IEEE ICC/GLOBECOM optical symposium at GLOBECOM'00, the Future Photonic Network Technologies, Architectures and Protocols Symposium. He chaired this Symposium, which continues to date under different names. He was the founding chair of the first Green Track at ICC/GLOBECOM at GLOBECOM 2011, and is Chair of the IEEE Sustainable ICT Initiative within the IEEE Technical Activities Board (TAB) Future Directions Committee (FDC) and within the IEEE Communications Society, a pan IEEE Societies Initiative responsible for Green and Sustainable ICT activities across IEEE, 2012-present. He is and has been on the technical program committee of 39 IEEE ICC/GLOBECOM conferences between 1995 and 2020 including 19 times as Symposium Chair. He received the IEEE Communications Society Hal Sobol award, the IEEE Comsoc Chapter Achievement award for excellence in chapter activities (both in 2005), the University of Wales Swansea Outstanding Research Achievement Award, 2006, the IEEE Communications Society Signal Processing and Communication Electronics outstanding service award, 2009, a best paper award at IEEE ICC'2013, the IEEE Comsoc Transmission Access and Optical Systems outstanding Service award 2015 in recognition of "Leadership and Contributions to the Area of Green Communications", received the GreenTouch 1000x award in 2015 for "pioneering research contributions to the field of energy efficiency in telecommunications", the 2016 IET Optoelectronics Premium Award and shared with 6 GreenTouch innovators the 2016 Edison Award in the "Collective Disruption" Category for their work on the GreenMeter, an international competition, clear evidence of his seminal contributions to Green Communications which have a lasting impact on the environment (green) and society. He is currently an editor of: IET Optoelectronics, Journal of Optical Communications and was editor of IEEE Communications Surveys and Tutorials and IEEE Journal on Selected Areas in Communications series on Green Communications and Networking. He was Co-Chair of the GreenTouch Wired, Core and Access Networks Working Group, an adviser to the Commonwealth Scholarship Commission, member of the Royal Society International Joint Projects Panel and member of the Engineering and Physical Sciences Research Council (EPSRC) College. He was Principal Investigator (PI) of the £6m EPSRC INTelligent Energy awaRe NETworks (INTERNET) Programme Grant, 2010-2016 and is currently PI of the £6.6m EPSRC Terabit Bidirectional Multi-user Optical Wireless System (TOWS) for 6G LiFi Programme Grant, 2019-2024. He has been awarded in excess of £30 million in grants to date from EPSRC, the EU and industry and has held prestigious fellowships funded by the Royal Society and by BT. He was an IEEE Comsoc Distinguished Lecturer 2013-2016.